\documentclass[aps,preprintnumbers,prl,twocolumn,superscriptaddress]{revtex4}

\usepackage{epsfig,latexsym,cancel,amssymb,amsmath}
\usepackage{graphicx}
\usepackage{feynmp}
\usepackage{subfigure}
\usepackage{epstopdf}
\usepackage{color}

\def\be{\begin{equation}}
\def\ee{\end{equation}}
\def\bea{\begin{eqnarray}}
\def\eea{\end{eqnarray}}

\def\bbuildrel#1_#2^#3{\mathrel{\mathop{\kern 0pt#1}\limits_{#2}^{#3}}}
\def\slash#1{\setbox0=\hbox{$#1$}#1\hskip-\wd0\dimen0=5pt\advance
       \dimen0 by-\ht0\advance\dimen0 by\dp0\lower0.5\dimen0\hbox
         to\wd0{\hss\sl/\/\hss}}

\newcommand{\gae}{\lower 2pt \hbox{$\, \buildrel {\scriptstyle >}\over {\scriptstyle
\sim}\,$}}
\newcommand{\lae}{\lower 2pt \hbox{$\, \buildrel {\scriptstyle <}\over {\scriptstyle
\sim}\,$}}

\newcommand{\beq}{\begin{eqnarray}}
\newcommand{\eeq}{\end{eqnarray}}

\newcommand{\ba}{\begin{array}}
\newcommand{\ea}{\end{array}}

\long\def\symbolfootnote[#1]#2{\begingroup%
\def\thefootnote{\fnsymbol{footnote}}\footnote[#1]{#2}\endgroup}

\def\lsim{\mathrel{\rlap{\lower4pt\hbox{\hskip1pt$\sim$}}
    \raise1pt\hbox{$<$}}}         
\def\gsim{\mathrel{\rlap{\lower4pt\hbox{\hskip1pt$\sim$}}
    \raise1pt\hbox{$>$}}}         

\def\lsim{\:\raisebox{-0.5ex}{$\stackrel{\textstyle<}{\sim}$}\:}
\def\gsim{\:\raisebox{-0.5ex}{$\stackrel{\textstyle>}{\sim}$}\:}

\def\issue(#1,#2,#3){{\bf #1}, #2 (#3)}
\def\opcit(#1){ {\em op. cit.}, #1}

\def\APP(#1,#2,#3){Acta Phys.\ Polon.\ \issue(#1,#2,#3)}
\def\ARNPS(#1,#2,#3){Ann.\ Rev.\ Nucl.\ Part.\ Sci.\ \issue(#1,#2,#3)}
\def\CPC(#1,#2,#3){Comp.\ Phys.\ Comm.\ \issue(#1,#2,#3)}
\def\CIP(#1,#2,#3){Comput.\ Phys.\ \issue(#1,#2,#3)}
\def\EPJC(#1,#2,#3){Eur.\ Phys.\ J.\ C\ \issue(#1,#2,#3)}
\def\EPJD(#1,#2,#3){Eur.\ Phys.\ J. Direct\ C\ \issue(#1,#2,#3)}
\def\IEEETNS(#1,#2,#3){IEEE Trans.\ Nucl.\ Sci.\ \issue(#1,#2,#3)}
\def\IJMP(#1,#2,#3){Int.\ J.\ Mod.\ Phys. \issue(#1,#2,#3)}
\def\JHEP(#1,#2,#3){J.\ High Energy Physics \issue(#1,#2,#3)}
\def\JPG(#1,#2,#3){J.\ Phys.\ G \issue(#1,#2,#3)}
\def\MPL(#1,#2,#3){Mod.\ Phys.\ Lett.\ \issue(#1,#2,#3)}
\def\NP(#1,#2,#3){Nucl.\ Phys.\ \issue(#1,#2,#3)}
\def\NIM(#1,#2,#3){Nucl.\ Instrum.\ Meth.\ \issue(#1,#2,#3)}
\def\PL(#1,#2,#3){Phys.\ Lett.\ \issue(#1,#2,#3)}
\def\PRD(#1,#2,#3){Phys.\ Rev.\ D \issue(#1,#2,#3)}
\def\PRL(#1,#2,#3){Phys.\ Rev.\ Lett.\ \issue(#1,#2,#3)}
\def\SJNP(#1,#2,#3){Sov.\ J. Nucl.\ Phys.\ \issue(#1,#2,#3)}
\def\ZPC(#1,#2,#3){Zeit.\ Phys.\ C \issue(#1,#2,#3)}

\def\beq{\begin{equation}}
\def\eeq{\end{equation}}
\def\bea{\begin{eqnarray}}
\def\eea{\end{eqnarray}}
\def\to{\rightarrow}

\begin{document}


\preprint{
\vbox{
\hbox{NPAC-11-16} 
}}

\title{Hadronic Light-by-Light and the Pion Polarizability}
\author{Kevin T. Engel}
\affiliation{California Institute of Technology, Pasadena, CA 91125 USA}
\author{Hiren H. Patel}
\affiliation{Department of Physics, University of Wisconsin, Madison, WI 53706, USA}
\author{Michael J. Ramsey-Musolf}
\affiliation{Department of Physics, University of Wisconsin, Madison, WI 53706, USA\\ and Kellogg Radiation Laboratory, California Institute of Technology, Pasadena, CA 91125}

\begin{abstract} We compute the charged pion loop contribution to the light-by-light scattering amplitude for off-shell photons in chiral perturbation theory through next-to-leading order (NLO). We show that NLO contributions are relatively more important due to a fortuitous numerical suppression of the leading-order (LO) terms. Consequently, one expects theoretical predictions for the hadronic light-by-light (HLBL) contribution to the muon anomalous magnetic moment, $a_\mu^\mathrm{HLBL}$, to be sensitive to the choice of model for the higher momentum-dependence of the LBL amplitude. We show that models employed thus far for the charged pion loop contribution to $a_\mu^\mathrm{HLBL}$ are not consistent with low-momentum behavior implied by quantum chromodynamics, having omitted potentially significant contributions from the pion polarizability. 

\end{abstract}

\maketitle

The anomalous magnetic moment of the muon, $a_\mu=(g_\mu-2)/2$,  continues to be a quantity of considerable interest in particle and nuclear physics. The present experimental value, $a_\mu^\mathrm{exp}= 116592089(63)\times 10^{-11}$ obtained by the E821 Collaboration\cite{Bennett:2006fi,Bennett:2004pv,Bennett:2002jb} differs from theoretical expectations by $3.6\sigma$ assuming the Standard Model (SM) of particle physics and state-of-the-art computations of hadronic contributions, including those obtained using data on  $\sigma(e^+e^-\to\mathrm{hadrons})$ and dispersion relation methods: $a_\mu^\mathrm{SM}= 116591802(49)\times 10^{-11}$ (for recent reviews, see Ref.~\cite{Passera:2010ev,Jegerlehner:2009ry} as well as references therein). A deviation of this magnitude can be naturally explained in a number of scenarios for physics beyond the Standard Model (BSM), including (but not limited to) supersymmetry, extra dimensions, or additional neutral gauge bosons \cite{Stockinger:2006zn,Hertzog:2007hz,Czarnecki:2001pv} . A next generation experiment planned for Fermilab would reduce the experimental uncertainty by a factor of four\cite{FERMILAB-PROPOSAL-0989}. If a corresponding reduction in the theoretical, SM uncertainty were achieved, the muon anomalous moment could provide an even more powerful indirect probe of BSM physics.

The most significant pieces of the error quoted above for $a_\mu^\mathrm{SM}$ are associated with the leading order hadronic vacuum polarization (HVP) and the HLBL contributions: $\delta a_\mu^\mathrm{HVP}(\mathrm{LO}) = \pm 42 \times 10^{-11}$ and $\delta a_\mu^\mathrm{HLBL} = \pm 26 \times 10^{-11}$ \cite{Prades:2009tw}  (other authors give somewhat different error estimates for the latter \cite{Hayakawa:1997rq,Hayakawa:1996ki,Hayakawa:1995ps,Knecht:2001qg,Knecht:2001qf,Melnikov:2003xd,Bijnens:2007pz,Nyffeler:2009tw,RamseyMusolf:2002cy} , but we will refer to these numbers as points of reference; see \cite{Nyffeler:2010rd} for a review). In recent years, considerable scrutiny has been applied to the determination of $a_\mu^\mathrm{HVP}(\mathrm{LO})$ from data on  $\sigma(e^+e^-\to\mathrm{hadrons})$ and hadronic $\tau$ decays. Use of the latter indicating a somewhat smaller discrepancy between the SM and experimental values for $a_\mu$ than quoted above. Clearly, a significant improvement in this determination will be needed if the levels of theoretical and future experimental precision are to be commensurate. 

Here, we concentrate on the $a_\mu^\mathrm{HLBL}$, focusing in particular on the contributions from charged pion loops. Subsequent to the first results from the E821 Collaboration, the theoretical community devoted substantial effort to refining the predictions for pseudoscalar \lq\lq pole" contributions, which appear at leading order in the expansion of the number of colors $N_C$ and which are numerically dominant. However, the error quoted for the charged pion loop contributions, which enter at subleading order in $N_C$, is now comparable to the uncertainty associated with the pseudoscalar pole terms. Thus, we are motivated to revisit the former as part of the effort to improve the level of confidence in the theoretical SM prediction for $a_\mu^\mathrm{HLBL}$.

As a first step in that direction, we have computed the HLBL scattering amplitude for off-shell photons to NLO in Chiral Perturbation Theory ($\chi$PT). $\chi$PT is an effective field theory for low-energy interactions of hadrons and photons that incorporates the approximate chiral symmetry of quantum chromodynamics (QCD) for light quarks. Long-distance hadronic effects can be computed order-by-order in an expansion of $p/\Lambda_\chi$, where $p$ is a typical energy scale (such as the pion mass $m_\pi$ or momentum) and $\Lambda_\chi=4\pi F_\pi\sim 1$ GeV is the hadronic scale with $F_\pi=93.4$ MeV being the pion decay constant. At each order in the expansion, presently incalculable strong interaction effects associated with energy scales of order $\Lambda_\chi$ are parameterized by a set of effective operators whose coefficients -- \lq\lq low energy constants" (LECs) -- are fit to experimental results and then used to predict other low-energy observables. 

$\chi$PT  has been applied with considerable success to the analysis of a variety of hadronic and electromagnetic processes (for a recent review, see {\em e.g.} \cite{Bijnens:2006zp}), making it an in principle appropriate and model-independent framework for investigating hadronic contributions to $a_\mu$, another low-energy observable. In the $\chi$PT analysis of the pseudoscalar pole contributions to $a_\mu^\mathrm{HLBL}$, however, one encounters a new LEC that cannot be determined independent of the $a_\mu$ measurement itself. Consequently, hadronic modeling is presently unavoidable if one wishes to predict the anomalous moment. Nevertheless, the calculable terms in $\chi$PT can be used to test or constrain model input, as any credible model for the LBL amplitude must reproduce behavior in the low-energy regime that is dictated by QCD. Indeed, the $\chi$PT computation of the leading $\ln^2$ term in the pion pole contribution revealed a critical sign error in earlier numerical computations of the pion pole contribution\cite{Knecht:2001qg,Knecht:2001qf}. The sub-leading $\ln$ term can be obtained from a combination of analytic computation\cite{RamseyMusolf:2002cy} and a determination of the relevant LEC from a determination of the $\pi^0\to e^+e^-$ branching ratio\cite{Abouzaid:2006kk}, and it can be used to further constrain the model input.

In this spirit, we have analyzed the charged pion loop contribution to the LBL amplitude to NLO and have compared with corresponding  predictions implied by models used in the computation of $a_\mu^\mathrm{HLBL}$. The leading order (in chiral counting) contribution is fixed entirely by gauge invariance and contains no unknown constants. As we show below, this contribution is fortuitously suppressed. As a result, higher order contributions are likely to be relatively more important than one might expect on general grounds, rendering this quantity more susceptible to model-dependent uncertainties.  Thus, it becomes all the more important that any model used for the charged pion contribution to $a_\mu^\mathrm{HLBL}$ respect the requirements of QCD at NLO in the low-momentum regime.  In this respect, we find that models utilized to date have omitted a potentially significant contribution associated with the pion polarizability, leading one to question the reliability of the presently-quoted value for $a_\mu^\mathrm{HLBL}$. Below, we provide details of the calculation leading to this conclusion. 

We compute the charged pion contributions to the LBL vertex function  $\Pi^{\mu\nu\alpha\beta}$ through NLO from the diagrams in Figure 1, expanding the result as a power series in the external (photon) momentum and pion mass. 
The LO amplitude that corresponds to a pure scalar QED calculation for point-like charged pions follows from Fig. 1(a) 
and yields a finite result that is free from any LECs. The result contains two $\mathcal{O}(p^4)$ structures that can be expressed in terms of two dimension eight ($d=8$) operators, $32\, \mathcal{O}_1^{(8)}\equiv(F^2)^2\equiv(F_{\mu\nu} F^{\mu\nu})^2$ and $8\,\mathcal{O}_2^{(8)}\equiv F^4=F_{\alpha\beta} F^{\beta\gamma} F_{\gamma\lambda} F^{\lambda\alpha}$, whose coefficients are given in Table \ref{tab:LO} (the operators are defined to absorb symmetry factors). Naively, one would expect the magnitude of the coefficients to be set by $1/(4\pi)^2 \times 1/m_\pi^4$ . However, we find that each operator contains an additional suppression factor of $1/9$ and $1/45$, respectively. Thus, we anticipate that the NLO contributions from the graphs of Fig. 1(b-d) will be relatively more important. 

\begin{figure}
\subfigure[$\,$LO $\pi^\pm$ contribution to LBL]{
\includegraphics[width=56mm,height=15mm]{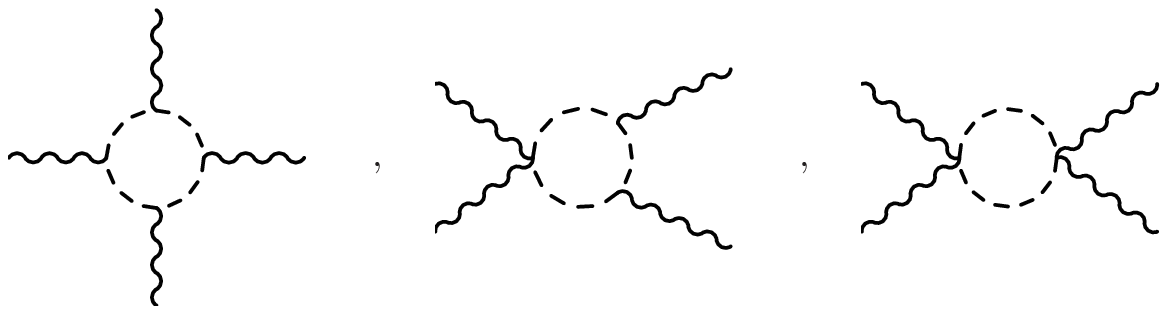}
}
\subfigure[$\,$NLO propagator correction]{
\includegraphics[width=40mm,height=15mm]{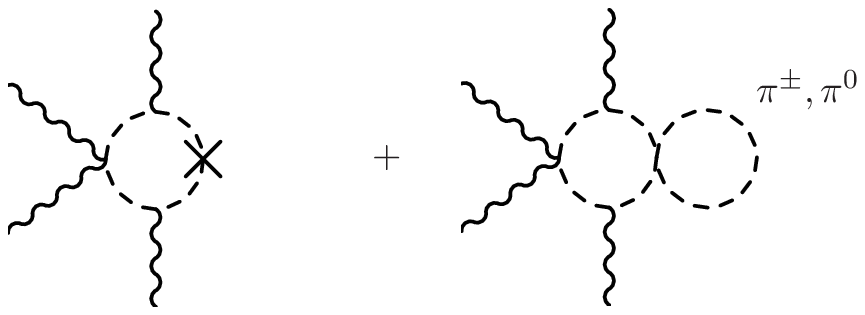}
}
\subfigure[$\,$NLO $\gamma\pi\pi$ vertex correction]{
\includegraphics[width=63mm,height=15mm]{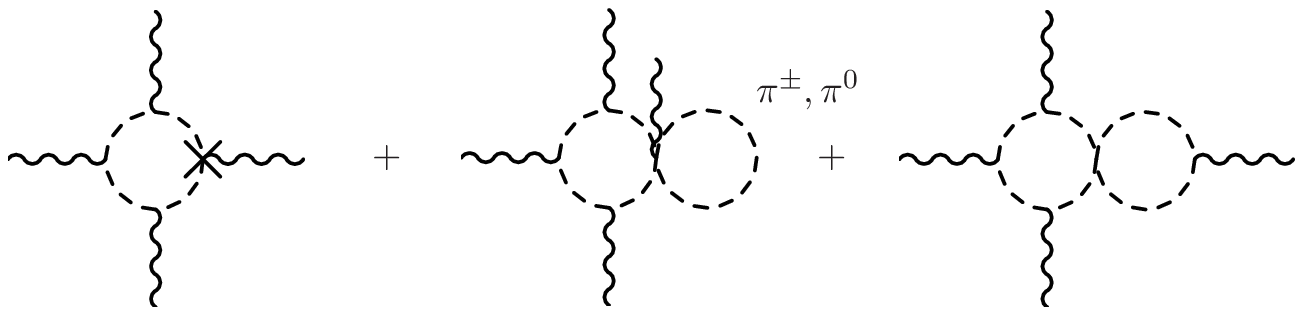}
}
\subfigure{
\includegraphics[width=72mm,height=11mm]{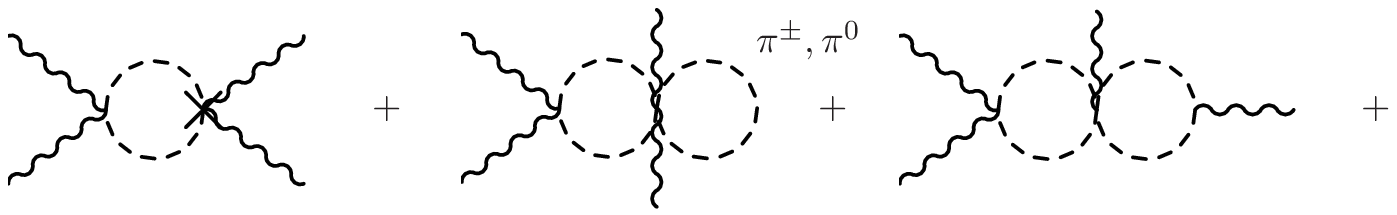}
}
\renewcommand{\thesubfigure}{(d)}
\subfigure[$\,$NLO $\gamma\gamma\pi\pi$ vertex correction]{
\includegraphics[width=47mm,height=9mm]{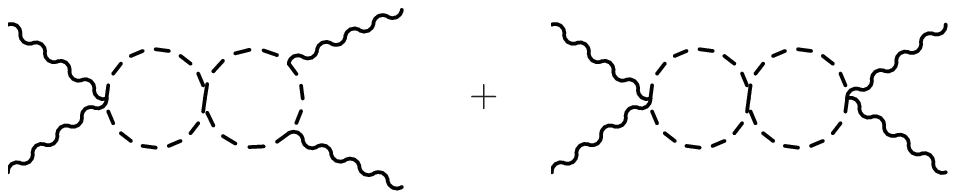}
}
\caption{Representative diagrams for charged pion loop contributions to the LO and NLO to LBL amplitude.}
\end{figure}

The graphs in Figures 1(b-d) correspond respectively to the propagator, vertex, and polarizability corrections. The first two classes are divergent and require the introduction of counterterms from the $\mathcal{O}(p^4)$ chiral Lagrangian. We carry out the calculation using dimensional regularization in $d=4-2\epsilon$ dimensions and define the counterterms to remove the contributions proportional to $1/\epsilon-\gamma+\ln 4\pi +1$ as is the standard convention for $\chi$PT\cite{Bijnens:2006zp}.  We find that the explicit dependence on the counterterms needed for renormalization of the pion propagator is cancelled by charge and mass renormalization, leaving only a dependence on the $\mathcal{O}(p^4)$ operator associated with the charge radius of the pion:
\be
\mathcal{L}_9 = ie \alpha_9\ F_{\mu\nu}\, \mathrm{Tr}\ \left(Q\left[D^\mu\Sigma, D^\nu\Sigma^\dag\right]\right)\ \ \ ,
\ee
where $Q=\mathrm{diag}(2/3,-1/3)$ is the electric charge matrix and $\Sigma=\mathrm{exp}(i \tau^a\ \pi^a/F_\pi)$ with $a=1,2,3$ giving the non-linear realization of the spontaneously broken chiral symmetry. After renormalization, one has for the square of the pion charge radius
\be
r_\pi^2 = \frac{12}{F_\pi^2}\alpha_9^r(\mu)+\frac{1}{\Lambda_\chi^2}\left[ \ln\left(\frac{\mu^2}{m_\pi^2}\right)-1\right]\ \ \ 
\ee
where the superscript \lq\lq $r$" indicates the finite component after the subtraction of $1/\epsilon-\gamma+\ln 4\pi+1$ term is performed. Choosing $\mu=m_\rho$ and taking the experimental value for $r_\pi^2$ gives $\alpha_9^r(m_\rho)= (7.0\pm0.2)\times 10^{-3}$  for two-flavor $\chi$PT at $\mathcal{O}(p^4)$. Within error bars, this result is the same as obtained in Ref.~\cite{hep-ph/0203049} for the three-flavor case. 

The $\pi\pi\gamma\gamma$ vertex correction shown in Fig. 1(d) is finite, but the polarizability amplitude nevertheless receives an additional finite contribution from $\mathcal{L}_9$ and 
\be
\mathcal{L}_{10} = e^2\alpha_{10}\ F^2\ \mathrm{Tr}\left(Q\Sigma Q\Sigma^\dag\right)\ \ \ .
\ee
The corresponding combination entering the LBL amplitude is $\alpha_9^r+\alpha_{10}^r$. As the sum of the one-loop polarizability sub-graphs is finite, this combination of LECs is independent of the renormalization scale. An experimental value $(\alpha_9^r+\alpha_{10}^r)_\mathrm{exp} = (1.32\pm 0.14)\times 10^{-3}$ has been obtained from radiative pion decay \cite{arXiv:0801.2482}. As a cross check on the extraction of these LECs we also consider the determination of $\alpha_{10}^r$ from semileptonic $\tau$-decays given in Ref.~\cite{arXiv:0810.0760}. Converting from three- to two-flavor $\chi$PT we obtain $\alpha_{10}^r(m_\rho)=-(5.19\pm 0.06)\times 10^{-3}$, in reasonable agreement with the determination of $\alpha_9^r(m_\rho)$ from the pion form factor and $(\alpha_9^r+\alpha_{10}^r)$ from pion radiative decay. The resulting prediction for the pion polarizability\cite{Gasser:2006qa}, which we confirm by taking the on-shell photon limit of our off-shell $\pi^+\pi^-\gamma\gamma$ computation, disagrees with the latest experimental determination\cite{Ahrens:2004mg} by a factor of two. 

The final NLO results for the LBL amplitude are summarized in Table I. To lowest order in external momenta, the only change from LO are polarizability corrections which modify the $\mathcal{O}^{(8)}_1$ coefficient. To see the full impact of the (higher momentum) NLO terms, we expand our result to $\mathcal{O}(p^6)$, introducing a complete basis of seven  $d=10$ four-photon operators:
\begin{eqnarray}\nonumber
16\, \mathcal{O}_1^{(10)} &=&  \partial_\rho F_{\mu\nu}\partial^\rho F^{\mu\nu}F_{\alpha\beta}F^{\alpha\beta} \\
\nonumber
8\  \mathcal{O}_2^{(10)}&=&\partial_\rho F_{\mu\nu}F^{\mu\nu}\partial^\rho F_{\alpha\beta}F^{\alpha\beta} \\
\nonumber
2\ \mathcal{O}_3^{(10)}& =&\partial_\rho F_{\alpha\beta}\partial^\rho F^{\beta\gamma}F_{\gamma\delta}F^{\delta\alpha}\\
\nonumber
4\  \mathcal{O}_4^{(10)}&=&\partial_\rho F_{\alpha\beta}F^{\beta\gamma}\partial^\rho F_{\gamma\delta}F^{\delta\alpha} \\
\nonumber
4\ \mathcal{O}_5^{(10)}&=&\partial^\mu F_{\mu\nu}F^{\alpha\nu}\partial_\alpha F_{\beta\gamma}F^{\beta\gamma} \\
\nonumber
4\  \mathcal{O}_6^{(10)}&=&F_{\mu\nu}F^{\alpha\nu}\partial^\mu F_{\beta\gamma}\partial_\alpha F^{\beta\gamma} \\
\nonumber
2\ \mathcal{O}_7^{(10)}&=& F_{\mu\nu}\partial^\mu F_{\alpha\beta}\partial^\nu F^{\beta\gamma}F_{\gamma\alpha} 
\end{eqnarray}
The coefficients of these operators are given in Table II.  At this order, both vertex and polarizability corrections modify the LO result.  

To obtain a sense of the numerical impact of the two-loop corrections, including those involving $\alpha_9^r+\alpha_{10}^r$, we utilize the values of the LECs discussed above. In the case of $\mathcal{O}_1^{(8)}$, the NLO (two-loop) contribution represents a $\sim 20\%$ correction to the LO term, substantially larger than the $\sim m_\pi^2/\Lambda_\chi^2\sim 0.01$ magnitude one might expect from power counting arguments. In the case of the $d=10$ operators, the NLO corrections range from a few to $\sim 30\%$. The largest impact of the charge radius corrections is on $\mathcal{O}_1^{(10)}$ ($\sim 30\%$)  while the most important effect of the polarizability is on $\mathcal{O}_2^{(10)}$ ($\sim 10\%$). As we discuss below, the numerical impact of the various NLO contributions on the low-momentum HLBL amplitude -- while illustrating their relative importance due to the LO suppression -- may not be indicative of their impact on the $a_\mu^\mathrm{HLBL}$. Indeed, previous experience with the inclusion of the pion form factor in earlier work  \cite{Hayakawa:1995ps,Bijnens:1995cc,Bijnens:1995xf,Hayakawa:1996ki} suggests that the effect on $a_\mu^\mathrm{HLBL}$ may be even more pronounced than implied by these low-momentum comparisons.

\begin{table}[hptb]
\caption{Coefficients of lowest dimension ($d=8$) operators contributing to the HLBL amplitude, scaled by $(4\pi)^2 m_\pi^4/e^4$. Second and third columns give LO and NLO contributions in $\chi$PT, while final column indicates the VMD result 
\cite{Bijnens:1995xf}.  
 } \label{tab:LO}
\begin{tabular}{| c | c | c | c |}
\hline
Operator & 1 loop $\chi$PT & 2 loop  & VMD \\ \hline
$\mathcal{O}_1^{(8)}$ & $1/9$ & 
$\frac{m_\pi^2}{F_\pi^2}\frac{16}{3}(\alpha_9^r+\alpha_{10}^r)$ & 0 \\ \hline
$\mathcal{O}_2^{(8)}$ & $1/45$ & 0 & 0 \\ \hline
\end{tabular}
\end{table}

\begin{table}[hptb]
\caption{Coefficients of $d=10$ operators $\mathcal{O}_n^{(10)}$  contributing to the HLBL amplitude, scaled by $(4\pi)^2 m_\pi^6/e^4$. First column denotes operator index $n$. Second and third columns give LO and NLO contributions in $\chi$PT, while final column indicates VMD result.  Identifying $r_\pi^2=6/M_V^2$ (see text) implies agreement between the two-loop $\chi$PT and VMD predictions for the charge radius contribution. 
 } \label{tab:NLO}
\begin{tabular}{| c | c | c | c |}
\hline
$n$ & 1 loop  & 2 loop  & VMD \\ \hline
$1$ &$\frac{1}{45}$ &$\frac{1}{3} \bigl\{\frac{1}{9}(m_\pi r_\pi)^2+\frac{4}{5}(\frac{m_\pi}{F_\pi})^2(\alpha_9^r+\alpha_{10}^r) \bigr\}$ & $ \frac{2}{9}\frac{m_\pi^2}{M_V^2}$ \\ \hline

$2$ & $\frac{2}{45}$ & $\frac{1}{9}\bigl\{\frac{1}{3} (m_\pi r_\pi)^2+\frac{1}{2}\frac{m_\pi^2}{\Lambda_\chi^2}+\frac{44}{5} (\frac{m_\pi}{F_\pi})^2 (\alpha_9^r+\alpha_{10}^r)\bigr\}$ & $\frac{2}{9}\frac{m_\pi^2}{M_V^2}$ \\ \hline

$3$ & $\frac{2}{315}$ & $ \frac{1}{135} (m_\pi r_\pi)^2 $ & $\frac{2}{45}\frac{m_\pi^2}{M_V^2}$ \\ \hline

$4$ & $\frac{1}{189}$ & $ \frac{1}{135} (m_\pi r_\pi)^2 $ & $\frac{2}{45}\frac{m_\pi^2}{M_V^2}$ \\ \hline

$5$ & $\frac{1}{135}$ & $\frac{4}{45} (\frac{m_\pi}{F_\pi})^2 (\alpha_9^r+\alpha_{10}^r)$ & 0 \\ \hline

$6$ & $\frac{1}{315}$ & 0 & 0 \\ \hline

$7$ & $\frac{1}{945}$ & 0 & 0\\ \hline

\end{tabular}
\end{table}

We now compare the explicit NLO results in $\chi$PT with the corresponding expectations for the operators in Tables I and II derived from models used to compute the charged pion loop contribution to   $a_\mu^\mathrm{HLBL}$. For concreteness, we focus on the extended Nambu-Jona-Lasinio (ENJL) model adopted in Ref.~\cite{Bijnens:1995xf}. In that work, the point-like contributions to the LBL vertex function $\Pi^{\mu\nu\alpha\beta}$ are modified by the inclusion of vector meson dominance (VMD) type propagator functions
$
V_{\mu\lambda}(k^2) = ( g_{\mu\lambda} M_V^2-p_\mu p_\lambda)/( M_V^2-p^2)
$
as
\be
\label{eq:vmd1}
\Pi^{\mu\nu\alpha\beta}\to V_{\mu\lambda}(p_1) V_{\nu\sigma}(p_2) V_{\alpha\rho}(p_3) V_{\beta\eta}(p_4)\ \Pi^{\lambda\sigma\rho\eta}\ \ \ ,
\ee
with the \lq\lq vector meson mass" $M_V$ in general a function of the photon momentum $p_j^2$. The Ward identities imply that the $p_\mu p_\lambda$ terms do not contribute to the overall LBL vertex function; hence, the replacement of Eq.~(\ref{eq:vmd1}) is equivalent introducing a VMD form factor for each photon when $M_V$ is taken to be a constant. The corresponding prediction for the  charge radius is $(r_\pi^2)_\mathrm{VMD} = 6/{M_V^2}$. For $M_V=m_\rho$, one obtains a value for $r_\pi^2$ in good agreement with experiment. An analogous treatment using a Hidden Local Symmetry approach~ \cite{Hayakawa:1995ps,Hayakawa:1996ki} agrees with the ENJL prescription to $\mathcal{O}(p^6)$.

Expanding the right hand side of Eq.~(\ref{eq:vmd1}) to first order in $p^2/M_V^2$ we obtain the model prediction for the NLO operator coefficients given in the last column of Table I. Identifying $6/M_V^2$ with the corresponding quantity that gives the pion charge radius  ,
we observe that the VMD model reproduces some but not all of the physics that one expects at NLO for the LBL amplitude. In particular, the polarizability contributions to $\mathcal{O}_1^{(8)}$ as well as $\mathcal{O}_{1,2,5}^{(10)}$ are absent from the VMD prescription. As a point of principle, the results of this comparison imply that the VMD-type models employed for $a_\mu^\mathrm{HLBL}$ are not fully consistent with the strictures of QCD for the low-momentum behavior of  $\Pi^{\mu\nu\alpha\beta}$ and that use of a  more consistent model prescription is warranted. 

On a practical level, given the relative magnitudes of the $\alpha_9^r+\alpha_{10}^r$ and $\alpha_9^r$, one has reason to suspect that the omission of the polarizability contribution could have numerically significant implications for  $a_\mu^\mathrm{HLBL}$. As discussed earlier, a comparison of the low-momentum LO and NLO contributions to the low-momentum HLBL amplitude indicates that the both the charge radius and polarizability contributions that appear at NLO can generate substantially larger corrections than one might expect based on power counting, due to the fortuitous numerical suppression of the LO terms. Moreover, the charge radius and polarizability contributions can have comparable magnitudes in the case of some operators, while for others, one or the other dominates. 

At this point, one may only speculate as to the effect on $a_\mu^\mathrm{HLBL}$ of the previously neglected polarizability contribution. Nevertheless, it is instructive to refer to existing model computations that introduce a pion form factor at the $\pi^+\pi^-\gamma$ vertices. In the original computation of Ref.~\cite{CLNS-84-606}, inclusion of the form factor via a VMD prescription reduced the magnitude of the charged pion loop contribution to $a_\mu^\mathrm{HLBL}$ by a factor of three from the scalar QED/point-like pion result. The subsequent computation using the HLS procedure yielded an even stronger suppression (a factor of ten)\cite{Hayakawa:1995ps,Hayakawa:1996ki}. The ENJL calculation of Ref.~\cite{Bijnens:1995xf} leads to a result that is about four times larger than the HLS computation, but still strongly suppressed compared to the point-like pion/scalar QED limit. In all cases, the use of a VMD type procedure that matches onto the $r_\pi^2$ terms for the HLBL amplitude at low-momentum has a much more significant  numerical impact on $a_\mu^\mathrm{HLBL}$ than the low-momentum comparisons would suggest. Given that the latter already indicate a substantial contribution from the pion polarizability, it appears important to include the corresponding physics in modeling the charged pion contribution to $a_\mu^\mathrm{HLBL}$. An effort to do so will be reported in forthcoming work.

\noindent \noindent{\it  Acknowledgements} 
We thank J. Bijnens, E. de Rafael, and A. Vainshtein for useful conversations and Mark B. Wise for discussions in the formulation of this project. 
The work is partially supported by U.S. Department of Energy contracts DE-NNN (KTE) and DE-FG02-08ER41531 (HHP and MJRM) and the Wisconsin Alumni Research Foundation (HHP and MJRM).


\end{document}